\documentclass[epj]{svjour}
\usepackage{indentfirst}
\usepackage{psfig,color}
\usepackage{epsfig}
\usepackage{epsf}
\usepackage{graphicx}

\usepackage{graphics}
\begin{document}
\title{Relations between quark and lepton mixing angles and
matrices}

\author{Nan Li \inst{1} \and Bo-Qiang Ma\inst{2,1,}\thanks{Email address: mabq@phy.pku.edu.cn; corresponding author. }}

%
%
\institute{Department of Physics, Peking University, Beijing
100871, China  \and CCAST (World Laboratory), P.O.~Box 8730,
Beijing 100080, China}

\date{Received: date / Revised version: date}
%
\abstract{We discuss the relations between the mixing angles and
the mixing matrices of quarks and leptons. With Raidal's numerical
relations, we parameterize the lepton mixing (PMNS) matrix with
the parameters of the quark mixing (CKM) matrix, and calculate the
products of $V_{\mathrm{CKM}}U_{\mathrm{PMNS}}$ and
$U_{\mathrm{PMNS}}V_{\mathrm{CKM}}$. Also, under the conjectures
$V_{\mathrm{CKM}}U_{\mathrm{PMNS}}=U_{\mathrm{bimax}}$ or
$U_{\mathrm{PMNS}}V_{\mathrm{CKM}}=U_{\mathrm{bimax}}$, we get the
PMNS matrix naturally, and test Raidal's relations in these two
different versions. The similarities and the differences between
the different versions are discussed in detail.
} 

\maketitle

\begin{onecolumn}

\section{\label{sec:level1}Introduction}

The mixing of quarks and leptons is one of the fundamental problem
in particle physics. But its origin is still unknown yet, and the
mixing is described phenomenologically by the mixing matrices,
i.e., the Cabibbo-Kobayashi-Maskawa (CKM)~\cite{ckm} matrix for
quark mixing and the Pontecorvo-Maki-Nakagawa-Sakata
(PMNS)~\cite{pmns} matrix for lepton mixing. To understand the
mixing problem, two aspects should be considered. One is the
mixing matrix, and the other is the mixing angle. However, these
mixing angles can not be determined by the standard model (SM)
itself, and can only be fixed by the experimental data. So the
mixing angles are taken as free parameters, and are not correlated
with each other. Furthermore, the quark and lepton mixing
matrices, which are composed of the mixing angles, are also
independent of each other. If we can find the relation between
these mixing angles or the relation between the mixing matrices,
it will be helpful for our understanding of the inner essence of
the SM and for the model construction of the grand unified theory.

In this paper, we discuss the relations between the mixing angles
and the mixing matrices of quarks and leptons, respectively.
First, for the mixing angles, Raidal has suggested some numerical
relations~\cite{raidal}
\begin{eqnarray}
&&\theta^{\mathrm{CKM}}_{1}+\theta_{1}^{\mathrm{PMNS}}(\theta_{\mathrm{atm}})=\frac{\pi}{4},\nonumber \\
&&\theta^{\mathrm{CKM}}_{2} \sim \theta_{2}^{\mathrm{PMNS}}(\theta_{\mathrm{chz}}) \sim \mathcal{O}(\lambda^3),\nonumber\\
&&\theta^{\mathrm{CKM}}_{3}(\theta_{\mathrm{C}})+\theta_{3}^{\mathrm{PMNS}}
(\theta_{\mathrm{sol}})=\frac{\pi}{4}, \label{eq.raidal}
\end{eqnarray}
where $\theta_i$ are the mixing angles of the CKM and the PMNS
matrices. With these relations, we can link the elements of the
CKM and the PMNS matrices together, and then can express the CKM
and the PMNS matrices in a unified way~\cite{Li}. Furthermore, we
can find the relation between these two mixing matrices.

Second, for the mixing matrices, we discuss the products of the
CKM and the PMNS matrices. Both
$V_{\mathrm{CKM}}U_{\mathrm{PMNS}}$ and
$U_{\mathrm{PMNS}}V_{\mathrm{CKM}}$ are calculated in detail. We
find that the product of the CKM and the PMNS matrices is rather
near the bimaximal mixing pattern. So we can get the PMNS matrix
in terms of the CKM matrix and the bimaximal mixing matrix. The
PMNS matrix can be parameterized by the parameters of the CKM
matrix, and the relations between the mixing angles are deduced
naturally.

In Sects.~2 and 3, we discuss the quark and lepton mixing
matrices, and the mixing angles and the parameterizations of quark
and lepton mixing matrices, and show their relations. In Sect.~4,
with the numerical relations 
between the quark and lepton mixing angles, we discuss the
relation between the quark and lepton mixing matrices, and point
out the similarities and the differences of different versions. In
Sect.~5, we discuss the relations between the mixing angles under
the conjecture that the product of quark and lepton mixing
matrices is the bimaximal mixing pattern. Some conclusions are
given in Sect.~6.


\section{The quark and lepton mixing matrices}

To see the generation of the quark mixing matrix, let us consider
the charge-changing weak current
\begin{equation}
j=2\sum\limits_{\alpha^{\prime}=u^{\prime},c^{\prime},t^{\prime}}\overline{u}_{\alpha^{\prime}}\gamma_{\rho}d_{\alpha^{\prime}},
\label{eq.j}
\end{equation}
where the u-type and d-type quark fields $u_{\alpha^{\prime}}$ and
$d_{\alpha^{\prime}}$ do not have definite masses, but are the
linear combinations of the massive quark fields $u_{\alpha}$ and
$d_{\alpha}$,
\begin{eqnarray}
u_{\alpha^{\prime}}=\sum\limits_{\alpha=u,c,t}V_u^{\alpha^{\prime}\alpha}u_{\alpha},
\quad
d_{\alpha^{\prime}}=\sum\limits_{\alpha=d,s,b}V_d^{\alpha^{\prime}\beta}d_{\beta},\label{eq.ud}
\end{eqnarray}
where $V_u$ and $V_d$ are unitary matrices, which can diagonalize
the quark mass matrices. Substituting Eq.~(\ref{eq.ud}) into
Eq.~(\ref{eq.j}), we have
\begin{eqnarray}
j&=&2\sum\limits_{\alpha^{\prime},\alpha,\beta}\overline{u}_{\alpha}\gamma_{\rho}V_{u}^{\alpha\alpha^{\prime}\dag}V_d^{\alpha^{\prime}\beta}d_{\beta}\nonumber\\
&=&2\sum\limits_{\alpha,\beta}\overline{u}_{\alpha}\gamma_{\rho}V_{\mathrm{CKM}}^{\alpha\beta}d_{\beta},\nonumber
\end{eqnarray}
where
\begin{equation}
V_{\mathrm{CKM}}=V_{u}^{\dag}V_d.\label{eq.ckm}
\end{equation}
$V_{\mathrm{CKM}}$ is the quark mixing (CKM) matrix, which links
the flavor eigenstates to the mass eigenstates of quarks.

The CKM matrix is measured by different experiments to a good
degree of accuracy~\cite{pdg}, and the elements of the modulus of
the CKM matrix are summarized as
\begin{eqnarray}\nonumber
\left(
      \begin{array}{ccc}
      0.9739-0.9751 & 0.221-0.227 & 0.0029-0.0045\\
      0.221-0.227 & 0.9730-0.9744 & 0.039-0.044\\
      0.0048-0.014 & 0.037-0.043 & 0.9990-0.9992\\
      \end{array}
        \right).\label{eq.modulusquark}
\end{eqnarray}
We can see that the CKM matrix is very near the unit matrix, and
it can be parameterized by the Wolfenstein
parametrization~\cite{wol}
\begin{equation}
   V_{\mathrm{CKM}}=\left(
        \begin{array}{ccc}
             1-\frac{1}{2}\lambda^2 & \lambda & A\lambda^3(\rho-i\eta)\\
             -\lambda & 1-\frac{1}{2}\lambda^2 & A\lambda^2\\
             A\lambda^3(1-\rho-i\eta) & -A\lambda^2 & 1\\
        \end{array}
        \right),\label{eq.wol}
\end{equation}
where $\lambda$ measures the strength of the deviation of
$V_{\mathrm{CKM}}$ from the unit matrix
($\lambda=\sin\theta_{\mathrm{C}}=0.2243\pm0.0016$,
$\theta_{\mathrm{C}}$ is the Cabibbo mixing angle), and $A$,
$\rho$ and $\eta$ are the other three parameters, with the best
fit values $A=0.82$, $\rho=0.20$ and $\eta=0.33$~\cite{pdg}.

Similarly, the lepton mixing (PMNS) matrix can be written as
\begin{equation}
U_{\mathrm{PMNS}}=U_{l}^{\dag}U_{\nu}.\label{eq.pmns}
\end{equation}
where $U_{l}$ and $U_{\nu}$ are unitary matrices, which can
diagonalize the charged-lepton and the neutrino mass matrices, and
$U_{\mathrm{PMNS}}$ links the flavor eigenstates to the mass
eigenstates of leptons.

The elements of the modulus of the PMNS matrix are summarized
as~\cite{garcia}
\begin{equation}\nonumber
\left(
        \begin{array}{ccc}
             0.77-0.88 & 0.47-0.61 & <0.20\\
             0.19-0.52 & 0.42-0.73 & 0.58-0.82\\
             0.20-0.53 & 0.44-0.74 & 0.56-0.81\\
        \end{array}
        \right).\label{eq.moduluslepton}
\end{equation}
We can see that the PMNS matrix deviates from the unit matrix very
much, but is quite near the bimaximal mixing pattern, which reads
\begin{equation}
          U_{\mathrm{bimax}}=
\left(
        \begin{array}{ccc}
            {\sqrt{2}}/{2} & {\sqrt{2}}/{2} & 0 \\
            -1/2 & 1/2 & {\sqrt{2}}/{2} \\
            1/2 & -1/2 & {\sqrt{2}}/{2}
        \end{array} \right).\label{eq.bimaximal}\nonumber
\end{equation}

Since the CKM matrix is quite near the unit matrix, and the PMNS
matrix is quite near the bimaximal matrix, we may assume that the
deviation of the PMNS matrix from the bimaximal can just be
described by the CKM matrix, that is
\begin{eqnarray}
U_{\mathrm{PMNS}}V_{\mathrm{CKM}}=U_{\mathrm{bimax}},\label{eq.uvb}
\end{eqnarray}
or
\begin{eqnarray}
V_{\mathrm{CKM}}U_{\mathrm{PMNS}}=U_{\mathrm{bimax}}.\label{eq.vub}
\end{eqnarray}
So we can get
\begin{eqnarray}
U_{\mathrm{PMNS}}=U_{\mathrm{bimax}}V_{\mathrm{CKM}}^{\dag},\label{eq.ubvdag}
\end{eqnarray}
or
\begin{eqnarray}
U_{\mathrm{PMNS}}=V_{\mathrm{CKM}}^{\dag}U_{\mathrm{bimax}}.\label{eq.uvdagb}
\end{eqnarray}

Eq.~(\ref{eq.uvb}) and Eq.~(\ref{eq.vub}) have both been pointed
out by Minakata and Smirnov~\cite{qlc}, and the similar results
have also been discussed in the literature~\cite{literature}.
Thus, the PMNS matrix can be expressed thoroughly by the CKM
matrix, and can be parameterized by the Wolfenstein parameters of
the CKM matrix. So we can get the relations between the mixing
angles of quarks and leptons. We will discuss these two cases in
Sec.~5.

\section{The mixing angles of the quark and lepton mixing
matrices}

Both the CKM matrix and the PMNS matrix can be written as
\begin{equation}
    \left(
        \begin{array}{ccc}
            c_{2}c_{3} & c_{2}s_{3} & s_{2}e^{-i\delta}\\
           -c_{1}s_{3}-s_{1}s_{2}c_{3}e^{i\delta} & c_{1}c_{3}-s_{1}s_{2}s_{3}e^{i\delta} & s_{1}c_{2}\\
            s_{1}s_{3}-c_{1}s_{2}c_{3}e^{i\delta} & -s_{1}c_{3}-c_{1}s_{2}s_{3}e^{i\delta} & c_{1}c_{2}\\
        \end{array}
        \right),\label{eq.mixing matrix}
\end{equation}
where $s_{i}=\sin\theta_{i}$, $c_{i}=\cos\theta_{i}$ (for $i=1, 2,
3$), which describe the mixings between 2nd and 3rd, 3rd and 1st,
and 1st and 2nd generations of quarks or leptons, and $\delta$ is
the Dirac $CP$-violating phase. Altogether there are eight (four
for quark sector and four for lepton sector) parameters in the
mixing matrices, describing both the real and the imaginary parts
of the mixing matrices. If neutrinos are of Majorana type, it is
always feasible to parameterize the neutrino mixing matrix as a
product of Eq.~(\ref{eq.mixing matrix}) and a diagonal phase
matrix with two unremovable phase angles $\mbox{diag} (1,
e^{i\alpha}, e^{i\beta})$~\cite{sch}, where $\alpha$, $\beta$ are
the Majorana CP-violating phases.

For quark sector, these angles are measured to a good degree of
accuracy (for example, see~\cite{pdg}). The best fit values of the
three mixing angles are $\theta_1^{\mathrm{CKM}}=2.4^{\circ}$,
$\theta_2^{\mathrm{CKM}}=0.2^{\circ}$, and
$\theta_3^{\mathrm{CKM}} (\theta_{\mathrm{C}}) =12.9^{\circ}$.

For lepton sector, with the help of various experimental data from
KamLAND~\cite{Kam}, SNO~\cite{sno}, K2K~\cite{K2K},
Super-Kamiokande~\cite{SUPER} and CHOOZ~\cite{Chz} experiments, we
now have a much better understanding of these mixing angles,
\begin{eqnarray}
&&\sin^{2}2\theta_{\mathrm{atm}}=1.00\pm0.05, \nonumber\\
&&\sin^{2}2\theta_{\mathrm{chz}}=0\pm0.065, \nonumber\\
&&\tan^2\theta_{\mathrm{sol}}=0.41\pm0.05,\label{eq.mixing
angles}\nonumber
\end{eqnarray}
where $\theta_{\mathrm{atm}}$, $\theta_{\mathrm{chz}}$, and
$\theta_{\mathrm{sol}}$ are the mixing angles of atmospheric,
CHOOZ and solar neutrino oscillations, and we have
$\theta_{\mathrm{atm}}=\theta_{1}^{\mathrm{PMNS}}=45.0^{\circ}\pm6.5^{\circ}$,
$\theta_{\mathrm{chz}}=\theta_{2}^{\mathrm{PMNS}}=0^{\circ}\pm7.4^{\circ}$
and
$\theta_{\mathrm{sol}}=\theta_{3}^{\mathrm{PMNS}}=32.6^{\circ}\pm1.6^{\circ}$~\cite{raidal}.

An interesting numerical relation between the third mixing angles
of quarks and leptons was pointed out by Smirnov~\cite{Smirnov},
\begin{equation}
\theta_{3}^{\mathrm{CKM}}(\theta_{\mathrm{C}})+\theta_{3}^{\mathrm{PMNS}}(\theta_{\mathrm{sol}})=\frac{\pi}{4}.\label{eq.smirnov}
\end{equation}
And this relation is called the quark-lepton complementarity
(QLC)~\cite{qlc}.

Raidal extended this relation to three generations~\cite{raidal},
\begin{eqnarray}
&&\theta^{\mathrm{CKM}}_{1}+\theta_{1}^{\mathrm{PMNS}}(\theta_{\mathrm{atm}})=\frac{\pi}{4},\nonumber \\
&&\theta^{\mathrm{CKM}}_{2} \sim \theta_{2}^{\mathrm{PMNS}}(\theta_{\mathrm{chz}}) \sim \mathcal{O}(\lambda^3),\nonumber\\
&&\theta^{\mathrm{CKM}}_{3}(\theta_{\mathrm{C}})+\theta_{3}^{\mathrm{PMNS}}
(\theta_{\mathrm{sol}})=\frac{\pi}{4}.\nonumber
\end{eqnarray}

With these relations, we can find that the mixing angles of quarks
and leptons are not independent of each other. And thus we can get
the trigonometric functions of the mixing angles of leptons in
terms of those of quarks, and link the parameters of the PMNS
matrix with those of the CKM matrix. Therefore, we can
parameterize the PMNS matrix and the CKM matrix in a same
framework~\cite{Li}. And then we can test the product relations in
Eq.~(\ref{eq.uvb}) and Eq.~(\ref{eq.vub}). We will discuss these
cases in Sec.~4.

\section{The relations between the mixing angles}

In Wolfenstein parametrization of the CKM matrix, we have (to the
order of $\lambda^3$)
\begin{eqnarray}\nonumber
&&\sin\theta^{\mathrm{CKM}}_{1}=A\lambda^2, \quad \cos\theta^{\mathrm{CKM}}_{1}=1,\nonumber\\
&&\sin\theta^{\mathrm{CKM}}_{2}e^{-i\delta}=A\lambda^3(\rho-i\eta), \quad \cos\theta^{\mathrm{CKM}}_{2}=1,\nonumber\\
&&\sin\theta^{\mathrm{CKM}}_{3}=\lambda,\quad
\cos\theta^{\mathrm{CKM}}_{3}=1-\frac{1}{2}\lambda^2.
\label{eq.wolfenstein}
\end{eqnarray}

Using Eq.~(\ref{eq.raidal}), we can get the trigonometric
functions of the mixing angles of leptons (to the order of
$\lambda^3$)
\begin{eqnarray}
&&\sin\theta_1^{\mathrm{PMNS}}=\sin(\frac{\pi}{4}-\theta^{\mathrm{CKM}}_{1})=\frac{\sqrt{2}}{2}(1-A\lambda^2), \nonumber \\
&&\cos\theta_1^{\mathrm{PMNS}}=\frac{\sqrt{2}}{2}(1+A\lambda^2),\nonumber\\
&&\sin\theta_2^{\mathrm{PMNS}}e^{-i\delta}=A\lambda^3(\zeta-i\xi),\nonumber\\
&&\cos\theta_2^{\mathrm{PMNS}}=1, \nonumber\\
&&\sin\theta_3^{\mathrm{PMNS}}=\frac{\sqrt{2}}{2}(1-\lambda-\frac{1}{2}\lambda^2), \nonumber\\
&&\cos\theta_3^{\mathrm{PMNS}}=\frac{\sqrt{2}}{2}(1+\lambda-\frac{1}{2}\lambda^2),
\label{eq.sincos}
\end{eqnarray}
where $A$ and $\lambda$ are the Wolfenstein parameters of the CKM
matrix. So the CKM and the PMNS matrices have only one set of
parameters with Raidal's numerical relations. Because there are in
total four angles in the mixing matrix (three mixing angles and
one $CP$-violating phase angle), and only two precise numerical
relations are known, we have to introduce another two new
parameters $\zeta$ and $\xi$ to describe the PMNS matrix fully.

In Eq.~(\ref{eq.sincos}), we set
$\sin\theta_2^{\mathrm{PMNS}}e^{-i\delta}=A\lambda^3(\zeta-i\xi)$.
Because of the inaccurate experimental data of neutrino
oscillations, we have not fixed the value of
$|U_{e3}^{\mathrm{PMNS}}|$, and only known its upper
bound~\cite{garcia}. Therefore, we may also set
$\sin\theta_2^{\mathrm{PMNS}}e^{-i\delta}=A\lambda^2(\zeta-i\xi)$.
Choosing which of them is to be determined by the future
experimental data, and we discuss these two cases here,
respectively.

{\it Case 1}:
$\sin\theta_2^{\mathrm{PMNS}}e^{-i\delta}=A\lambda^3(\zeta-i\xi)$.

Substituting Eq.~(\ref{eq.sincos}) into Eq.~(\ref{eq.mixing
matrix}), we can get the PMNS matrix as

\begin{eqnarray}\nonumber
   U_{\mathrm{PMNS}}&=&\left(
        \begin{array}{ccc}
            \frac{\sqrt{2}}{2}(1+\lambda-\frac{1}{2}\lambda^2) & \frac{\sqrt{2}}{2}(1-\lambda-\frac{1}{2}\lambda^2) & A\lambda^3(\zeta-i\xi) \\
            -\frac{1}{2}[1-\lambda+(A-\frac{1}{2})\lambda^2-A\lambda^3(1-\zeta-i\xi)] & \frac{1}{2}[1+\lambda+(A-\frac{1}{2})\lambda^2+A\lambda^3(1-\zeta-i\xi)] & \frac{\sqrt{2}}{2}(1-A\lambda^2) \\
            \frac{1}{2}[1-\lambda-(A+\frac{1}{2})\lambda^2+A\lambda^3(1-\zeta-i\xi)] & -\frac{1}{2}[1+\lambda-(A+\frac{1}{2})\lambda^2-A\lambda^3(1-\zeta-i\xi)] & \frac{\sqrt{2}}{2}(1+A\lambda^2)
        \end{array} \right)\\&=&
\left(
        \begin{array}{ccc}
            \frac{\sqrt{2}}{2} & \frac{\sqrt{2}}{2} & 0 \\
            -\frac{1}{2} & \frac{1}{2} & \frac{\sqrt{2}}{2} \\
            \frac{1}{2} & -\frac{1}{2} & \frac{\sqrt{2}}{2}
        \end{array} \right)+\lambda
\left(
        \begin{array}{ccc}
             \frac{\sqrt{2}}{2} & -\frac{\sqrt{2}}{2} & 0 \\
            \frac{1}{2} & \frac{1}{2} & 0\\
           - \frac{1}{2} & -\frac{1}{2} & 0
        \end{array} \right)+\lambda^{2}
\left(
        \begin{array}{ccc}
            -\frac{\sqrt{2}}{4} & -\frac{\sqrt{2}}{4} & 0 \\
            -\frac{1}{2}(A-\frac{1}{2}) & \frac{1}{2}(A-\frac{1}{2}) & -\frac{\sqrt{2}}{2}A \\
            -\frac{1}{2}(A+\frac{1}{2}) & \frac{1}{2}(A+\frac{1}{2}) & \frac{\sqrt{2}}{2}A
        \end{array} \right)\nonumber\\&&+\lambda^{3}
\left(
        \begin{array}{ccc}
            0 & 0 & A(\zeta-i\xi) \\
            \frac{1}{2}A(1-\zeta-i\xi) & \frac{1}{2}A(1-\zeta-i\xi) & 0\\
            \frac{1}{2}A(1-\zeta-i\xi) & \frac{1}{2}A(1-\zeta-i\xi) & 0
        \end{array} \right)+\cdots.\label{eq.3ci}
\end{eqnarray}

We can see from Eq.~(\ref{eq.3ci}) the followings.

(1). The bimaximal mixing pattern is deduced naturally as the
leading-order approximation as long as we accept the numerical
relations in Eq.~(\ref{eq.raidal}).

(2). The leading and next-to-leading order terms are just the same
as the expressions in the expansion of the PMNS matrix around the
bimaximal mixing pattern by Rodejohann~\cite{Rodejohann} and
us~\cite{li}.

(3). The Wolfenstein parameter $\lambda$ can characterize both the
deviation of the CKM matrix from the unit matrix (see
Eq.~(\ref{eq.wol})), and the deviation of the PMNS matrix from the
exactly bimaximal mixing pattern (see the next-to-leading order
term in Eq.~(\ref{eq.3ci})).

Since these two different kinds of deviations are characterized by
only one parameter set, the product of the CKM matrix and the PMNS
matrix may just be the exactly bimaximal mixing matrix
(Eq.~(\ref{eq.uvb}) and Eq.~(\ref{eq.vub})). To see this clearly,
we discuss these two versions of product, respectively.

(i) $V_{\mathrm{CKM}}U_{\mathrm{PMNS}}$

From Eq.~(\ref{eq.3ci}) and Eq.~(\ref{eq.wol}), we have
\begin{eqnarray}\nonumber
   V_{\mathrm{CKM}}U_{\mathrm{PMNS}}&=&
\left(
        \begin{array}{ccc}
            \frac{\sqrt{2}}{2} & \frac{\sqrt{2}}{2} & 0 \\
            -\frac{1}{2} & \frac{1}{2} & \frac{\sqrt{2}}{2} \\
            \frac{1}{2} & -\frac{1}{2} & \frac{\sqrt{2}}{2}
        \end{array} \right)+\lambda
\left(
        \begin{array}{ccc}
             \frac{\sqrt{2}-1}{2} & -\frac{\sqrt{2}-1}{2} & \frac{\sqrt{2}}{2} \\
             -\frac{\sqrt{2}-1}{2} & -\frac{\sqrt{2}-1}{2} & 0\\
             -\frac{1}{2} & -\frac{1}{2} & 0
        \end{array} \right)+\lambda^{2}
\left(
        \begin{array}{ccc}
            -\frac{\sqrt{2}-1}{2} & -\frac{\sqrt{2}-1}{2} & 0 \\
            -\frac{\sqrt{2}-1}{2} & \frac{\sqrt{2}-1}{2} & -\frac{\sqrt{2}}{4} \\
            -\frac{1}{4} & \frac{1}{4} & 0
        \end{array} \right)\nonumber\\&&+\lambda^{3}
\left(
        \begin{array}{ccc}
            -\frac{\sqrt{2}-1}{4}-\frac{1}{2}A(1-\rho+i\eta) & \frac{\sqrt{2}-1}{4}+\frac{1}{2}A(1-\rho+i\eta) & A[(\zeta-i\xi)-\frac{\sqrt{2}}{2}(1-\rho+i\eta)] \\
            \frac{\sqrt{2}-1}{4}-\frac{1}{2}A(\zeta+i\xi) & \frac{\sqrt{2}-1}{4}-\frac{1}{2}A(\zeta+i\xi) & 0\\
            \frac{1}{2}A[\sqrt{2}(1-\rho+i\eta)-(\zeta+i\xi)] & \frac{1}{2}A[\sqrt{2}(1-\rho+i\eta)-(\zeta+i\xi)] & 0
        \end{array} \right)\nonumber\\&&+\cdots.\label{eq.vu3}
\end{eqnarray}

We can see from Eq.~(\ref{eq.vu3}) that the deviation of the
product of the CKM matrix and the PMNS matrix from the exactly
bimaximal mixing matrix is of order $\lambda$.

(ii) $U_{\mathrm{PMNS}}V_{\mathrm{CKM}}$

Similarly, we have
\begin{eqnarray}\nonumber
   U_{\mathrm{PMNS}}V_{\mathrm{CKM}}&=&
\left(
        \begin{array}{ccc}
            \frac{\sqrt{2}}{2} & \frac{\sqrt{2}}{2} & 0 \\
            -\frac{1}{2} & \frac{1}{2} & \frac{\sqrt{2}}{2} \\
            \frac{1}{2} & -\frac{1}{2} & \frac{\sqrt{2}}{2}
        \end{array} \right)+\lambda^2
\left(
        \begin{array}{ccc}
            0 & 0 & \frac{\sqrt{2}}{2}A \\
            -\frac{1}{2}A & -\frac{1}{2}(\sqrt{2}-1)A & -\frac{1}{2}(\sqrt{2}-1)A \\
            -\frac{1}{2}A & -\frac{1}{2}(\sqrt{2}-1)A & \frac{1}{2}(\sqrt{2}-1)A
        \end{array} \right)\nonumber\\&&+\lambda^{3}
\left(
        \begin{array}{ccc}
            0 & 0 & A[(\zeta-i\xi)-\frac{\sqrt{2}}{2}(1-\rho+i\eta)] \\
            \frac{1}{2}A[\sqrt{2}(1-\rho-i\eta)-(\zeta+i\xi)] & -\frac{1}{2}A(\zeta+i\xi) & \frac{1}{2}A(1-\rho+i\eta)\\
            \frac{1}{2}A[\sqrt{2}(1-\rho-i\eta)-(\zeta+i\xi)] & -\frac{1}{2}A(\zeta+i\xi) & -\frac{1}{2}A(1-\rho+i\eta)
        \end{array} \right)+\cdots.\label{eq.uv3}
\end{eqnarray}

We can see from Eq.~(\ref{eq.uv3}) that the deviation of the
product of the PMNS matrix and the CKM matrix from the exactly
bimaximal mixing matrix is smaller (to the order of $\lambda^2$)
than the former one. So the conjecture in Eq.~(\ref{eq.uvb}) is
better than the conjecture in Eq.~(\ref{eq.vub}).

{\it Case 2}:
$\sin\theta_2^{\mathrm{PMNS}}e^{-i\delta}=A\lambda^2(\zeta-i\xi)$.

Repeating the process, we get
\begin{eqnarray}\nonumber
   U_{\mathrm{PMNS}}&=&\left(
        \begin{array}{ccc}
            \frac{\sqrt{2}}{2}(1+\lambda-\frac{1}{2}\lambda^2) & \frac{\sqrt{2}}{2}(1-\lambda-\frac{1}{2}\lambda^2) & A\lambda^2(\zeta-i\xi) \\
            -\frac{1}{2}\{1-\lambda-[\frac{1}{2}-A(1+\zeta+i\xi)]\lambda^2\} & \frac{1}{2}\{1+\lambda-[\frac{1}{2}-A(1-\zeta-i\xi)]\lambda^2\} & \frac{\sqrt{2}}{2}(1-A\lambda^2) \\
            \frac{1}{2}\{1-\lambda-[\frac{1}{2}+A(1+\zeta+i\xi)]\lambda^2\} & -\frac{1}{2}\{1+\lambda-[\frac{1}{2}+A(1-\zeta-i\xi)]\lambda^2\} & \frac{\sqrt{2}}{2}(1+A\lambda^2)
        \end{array} \right)\\&=&
\left(
        \begin{array}{ccc}
            \frac{\sqrt{2}}{2} & \frac{\sqrt{2}}{2} & 0 \\
            -\frac{1}{2} & \frac{1}{2} & \frac{\sqrt{2}}{2} \\
            \frac{1}{2} & -\frac{1}{2} & \frac{\sqrt{2}}{2}
        \end{array} \right)+\lambda
 \left(
        \begin{array}{ccc}
             \frac{\sqrt{2}}{2} & -\frac{\sqrt{2}}{2} & 0 \\
            \frac{1}{2} & \frac{1}{2} & 0\\
           - \frac{1}{2} & -\frac{1}{2} & 0
        \end{array} \right)\nonumber\\&&+\lambda^{2}
 \left(
        \begin{array}{ccc}
            -\frac{\sqrt{2}}{4} & -\frac{\sqrt{2}}{4} & A(\zeta-i\xi) \\
            \frac{1}{2}[\frac{1}{2}-A(1+\zeta+i\xi)] & -\frac{1}{2}[\frac{1}{2}-A(1-\zeta-i\xi)] & -\frac{\sqrt{2}}{2}A\\
            -\frac{1}{2}[\frac{1}{2}+A(1+\zeta+i\xi)] & \frac{1}{2}[\frac{1}{2}+A(1-\zeta-i\xi)] & \frac{\sqrt{2}}{2}A
        \end{array} \right)+\cdots.\label{eq.2ci}
\end{eqnarray}

And similarly, we have

(i) $V_{\mathrm{CKM}}U_{\mathrm{PMNS}}$

\begin{eqnarray}\nonumber
   V_{\mathrm{CKM}}U_{\mathrm{PMNS}}&=&
\left(
        \begin{array}{ccc}
            \frac{\sqrt{2}}{2} & \frac{\sqrt{2}}{2} & 0 \\
            -\frac{1}{2} & \frac{1}{2} & \frac{\sqrt{2}}{2} \\
            \frac{1}{2} & -\frac{1}{2} & \frac{\sqrt{2}}{2}
        \end{array} \right)+\lambda
\left(
        \begin{array}{ccc}
             \frac{\sqrt{2}-1}{2} & -\frac{\sqrt{2}-1}{2} & \frac{\sqrt{2}}{2} \\
             -\frac{\sqrt{2}-1}{2} & -\frac{\sqrt{2}-1}{2} & 0\\
             -\frac{1}{2} & -\frac{1}{2} & 0
        \end{array} \right)\nonumber\\&&+\lambda^{2}
\left(
        \begin{array}{ccc}
            -\frac{\sqrt{2}-1}{2} & -\frac{\sqrt{2}-1}{2} & A(\zeta-i\xi) \\
            -\frac{1}{2}[\sqrt{2}-1+A(\zeta+i\xi)] & \frac{1}{2}[\sqrt{2}-1-A(\zeta+i\xi)] & -\frac{\sqrt{2}}{4} \\
            -\frac{1}{2}[\frac{1}{2}+A(\zeta+i\xi)] & \frac{1}{2}[\frac{1}{2}-A(\zeta+i\xi)] & 0
        \end{array} \right)+\cdots.\label{eq.vu2}
\end{eqnarray}

and

(ii) $U_{\mathrm{PMNS}}V_{\mathrm{CKM}}$

\begin{eqnarray}\nonumber
   U_{\mathrm{PMNS}}V_{\mathrm{CKM}}&=&
\left(
        \begin{array}{ccc}
            \frac{\sqrt{2}}{2} & \frac{\sqrt{2}}{2} & 0 \\
            -\frac{1}{2} & \frac{1}{2} & \frac{\sqrt{2}}{2} \\
            \frac{1}{2} & -\frac{1}{2} & \frac{\sqrt{2}}{2}
        \end{array} \right)+\lambda^2
\left(
        \begin{array}{ccc}
            0 & 0 & A(\frac{\sqrt{2}}{2}+\zeta-i\xi) \\
            -\frac{1}{2}A(1+\zeta+i\xi) & -\frac{1}{2}A(\sqrt{2}-1+\zeta+i\xi) & -\frac{1}{2}(\sqrt{2}-1)A \\
            -\frac{1}{2}A(1+\zeta+i\xi) & -\frac{1}{2}A(\sqrt{2}-1+\zeta+i\xi) & \frac{1}{2}(\sqrt{2}-1)A
        \end{array} \right)\nonumber\\&&+\cdots.\label{eq.uv2}
\end{eqnarray}

Again, we find that the deviation of
$U_{\mathrm{PMNS}}V_{\mathrm{CKM}}$ from the exactly bimaximal
mixing matrix is rather small (to the order of $\lambda^2$), and
that the deviation of $V_{\mathrm{CKM}}U_{\mathrm{PMNS}}$ from the
exactly bimaximal mixing matrix is larger (to the order of
$\lambda$). So the former conjecture in Eq.~(\ref{eq.uvb}) in
still better than the conjecture in Eq.~(\ref{eq.vub}).

In summary, in both the cases of
$\sin\theta_2^{\mathrm{PMNS}}e^{-i\delta}=A\lambda^3(\zeta-i\xi)$
and
$\sin\theta_2^{\mathrm{PMNS}}e^{-i\delta}=A\lambda^2(\zeta-i\xi)$,
the product of $U_{\mathrm{PMNS}}V_{\mathrm{CKM}}$ is nearer to
the exactly bimaximal mixing matrix than the product of
$V_{\mathrm{CKM}}U_{\mathrm{PMNS}}$.

\section{The relations between the mixing matrices}

In the previous deductive process, we admit Raidal's numerical
relations between the mixing angles of quarks and leptons
beforehand, and thus get the PMNS matrix in terms of the
Wolfenstein parameters of the CKM matrix. Then we calculate the
product of $V_{\mathrm{CKM}}U_{\mathrm{PMNS}}$ and
$U_{\mathrm{PMNS}}V_{\mathrm{CKM}}$, and compare their deviations
from the exactly bimaximal mixing matrix. However, with the
current experimental data, we can also make the conjectures
$U_{\mathrm{PMNS}}V_{\mathrm{CKM}}=U_{\mathrm{bimax}}$ or
$V_{\mathrm{CKM}}U_{\mathrm{PMNS}}=U_{\mathrm{bimax}}$ at first,
and then get the PMNS matrix straightforward. Thereafter we can
find whether Raidal's relations hold well under these conjectures.
We discuss the two different products, respectively. We have seen
from Sec.~4 that $U_{\mathrm{PMNS}}V_{\mathrm{CKM}}$ is closer to
the bimaximal mixing pattern (to the order of $\lambda^2$) than
$V_{\mathrm{CKM}}U_{\mathrm{PMNS}}$ (to the order of $\lambda$),
so this time we discuss the case
$U_{\mathrm{PMNS}}V_{\mathrm{CKM}}=U_{\mathrm{bimax}}$ first.

{\it Case 1}:
$U_{\mathrm{PMNS}}V_{\mathrm{CKM}}=U_{\mathrm{bimax}}$.

We suggest this product as a possibility for the relation between
the quark and lepton mixing matrices. Although we have no
theoretical fundamental for this suggestion, we can see that this
product is consistent with Eq.~(\ref{eq.uv3}) and
Eq.~(\ref{eq.uv2}) in Sec.~4. In the following deductive process,
we can see that if we assume
$U_{\mathrm{PMNS}}V_{\mathrm{CKM}}=U_{\mathrm{bimax}}$, the QLC
can be obtained directly and Raidal's relations can hold good, and
the parametrization of the PMNS matrix can be deduced naturally.

Because $V_{\mathrm{CKM}}$ is unitary, we can get
$U_{\mathrm{PMNS}}$ by multiplying $V_{\mathrm{CKM}}^{\dag}$ on
the right side of $U_{\mathrm{bimax}}$,
\begin{eqnarray}\nonumber
   U_{\mathrm{PMNS}}&=&U_{\mathrm{bimax}}V_{\mathrm{CKM}}^{\dag}\\&=&
\left(
        \begin{array}{ccc}
            \frac{\sqrt{2}}{2} & \frac{\sqrt{2}}{2} & 0 \\
            -\frac{1}{2} & \frac{1}{2} & \frac{\sqrt{2}}{2} \\
            \frac{1}{2} & -\frac{1}{2} & \frac{\sqrt{2}}{2}
        \end{array} \right)+\lambda
\left(
        \begin{array}{ccc}
             \frac{\sqrt{2}}{2} & -\frac{\sqrt{2}}{2} & 0 \\
            \frac{1}{2} & \frac{1}{2} & 0\\
           - \frac{1}{2} & -\frac{1}{2} & 0
        \end{array} \right)+\lambda^{2}
\left(
        \begin{array}{ccc}
            -\frac{\sqrt{2}}{4} & -\frac{\sqrt{2}}{4} & -\frac{\sqrt{2}}{2}A \\
            \frac{1}{4} & \frac{\sqrt{2}}{2}A-\frac{1}{4} & -\frac{1}{2}A \\
            -\frac{1}{4} & \frac{\sqrt{2}}{2}A+\frac{1}{4} & \frac{1}{2}A
        \end{array} \right)\nonumber\\&&+\lambda^{3}
\left(
        \begin{array}{ccc}
            0 & 0 & \frac{\sqrt{2}}{2}A(1-\rho+i\eta) \\
            \frac{\sqrt{2}}{2}A(\rho+i\eta) & 0 & -\frac{1}{2}A(1-\rho+i\eta)\\
            \frac{\sqrt{2}}{2}A(\rho+i\eta) & 0 & \frac{1}{2}A(1-\rho+i\eta)
        \end{array} \right)+\cdots.\label{eq.linans}
\end{eqnarray}

We can see that the leading and the next-to-leading terms in
Eq.~(\ref{eq.linans}) are just the same as those in
Eq.~(\ref{eq.3ci}) and Eq.~(\ref{eq.2ci}). This indicates that
Raidal's relations (Eq.~(\ref{eq.raidal})) and
$U_{\mathrm{PMNS}}V_{\mathrm{CKM}}=U_{\mathrm{bimax}}$ are in very
good consistency with each other.

To see this more clearly, we can calculate the trigonometric
functions of the mixing angles of the PMNS matrix, and then
calculate the sums of the corresponding angles of quarks and
leptons.

From Eq.~(\ref{eq.linans}), we have
\begin{eqnarray}\nonumber
c_2^{\mathrm{PMNS}}s_3^{\mathrm{PMNS}}=\frac{\sqrt{2}}{2}-\frac{\sqrt{2}}{2}\lambda-\frac{\sqrt{2}}{4}\lambda^{2},\nonumber\\
c_2^{\mathrm{PMNS}}c_3^{\mathrm{PMNS}}=\frac{\sqrt{2}}{2}+\frac{\sqrt{2}}{2}\lambda-\frac{\sqrt{2}}{4}\lambda^{2}.\label{eq.linanc3s3}
\end{eqnarray}
From Eq.~(\ref{eq.linanc3s3}) we have (to the order of
$\lambda^{3}$)
\begin{equation}\nonumber
\tan\theta_{3}^{\mathrm{PMNS}}=1-2\lambda+2\lambda^2-3\lambda^3.\label{eq.tan}
\end{equation}
Thus, we can get (to the order of $\lambda^{3}$)
\begin{eqnarray}\nonumber
s_3^{\mathrm{PMNS}}=\frac{\sqrt{2}}{2}-\frac{\sqrt{2}}{2}\lambda-\frac{\sqrt{2}}{4}\lambda^{2},\nonumber\\
c_3^{\mathrm{PMNS}}=\frac{\sqrt{2}}{2}+\frac{\sqrt{2}}{2}\lambda-\frac{\sqrt{2}}{4}\lambda^{2}.\label{eq.cs}
\end{eqnarray}
Similarly, we have
\begin{eqnarray}\nonumber
s_1^{\mathrm{PMNS}}=\frac{\sqrt{2}}{2}-A\lambda^2+A\lambda^{3},\nonumber\\
c_1^{\mathrm{PMNS}}=\frac{\sqrt{2}}{2}+A\lambda^2-A\lambda^{3}.\label{eq.cs1}
\end{eqnarray}
Also, we have
\begin{equation}\nonumber
s_2^{\mathrm{PMNS}}e^{-i\delta}=-\frac{\sqrt{2}}{2}A\lambda^2+\frac{\sqrt{2}}{2}(1-\rho+i\eta)A\lambda^3,
\end{equation}
and so
\begin{equation}\nonumber
|s_2^{\mathrm{PMNS}}|=\frac{\sqrt{2}}{2}A\lambda^2\sqrt{(\lambda-\lambda\rho-1)^2+(\lambda\eta)^2}.
\end{equation}
Substituting the best fit values of $A$, $\lambda$, $\rho$ and
$\eta$, we have
\begin{equation}
|s_2^{\mathrm{PMNS}}|=0.48\lambda^2,\label{eq.s2}
\end{equation}
and $c_2^{\mathrm{PMNS}}=1$ (to the order of $\lambda^{3}$).

Now we have got all the six the trigonometric functions of the
mixing angles of leptons, and we can calculate the sums of the
mixing angles of quarks and leptons.

Using Eq.~(\ref{eq.wolfenstein}) and Eq.~(\ref{eq.cs}), we have
\begin{eqnarray}\nonumber
\sin(\theta_3^{\mathrm{CKM}}+\theta_3^{\mathrm{PMNS}})&=&s_3^{\mathrm{CKM}}c_3^{\mathrm{PMNS}}+c_3^{\mathrm{CKM}}s_3^{\mathrm{PMNS}}\\
&=&\frac{\sqrt{2}}{2},\nonumber
\end{eqnarray}
and thus
\begin{eqnarray}
\theta_3^{\mathrm{CKM}}+\theta_3^{\mathrm{PMNS}}=\frac{\pi}{4}.\label{eq.linan1}
\end{eqnarray}
We can find that the QLC is satisfied precisely.

Similarly,
\begin{eqnarray}\nonumber
\sin(\theta_1^{\mathrm{CKM}}+\theta_1^{\mathrm{PMNS}})=\frac{\sqrt{2}}{2}+(\frac{\sqrt{2}}{2}-1)A\lambda^2+A\lambda^3,\nonumber
\end{eqnarray}
and thus
\begin{eqnarray}
\theta_1^{\mathrm{CKM}}+\theta_1^{\mathrm{PMNS}}=\frac{\pi}{4}-(\sqrt{2}-1)A\lambda^2+\sqrt{2}A\lambda^3.\label{eq.linan2}
\end{eqnarray}
So Raidal's relation violates a little (to the order of
$\lambda^2$).

Also, for $s_2^{\mathrm{PMNS}}$, we can find from
Eq.~(\ref{eq.s2}) that $s_2^{\mathrm{PMNS}} \sim \lambda^2$, this
differs from Raidal's relation slightly, and is consistent with
the parametrization in Eq.~(\ref{eq.2ci}).

In summary, if we assume that
$U_{\mathrm{PMNS}}V_{\mathrm{CKM}}=U_{\mathrm{bimax}}$, we can get
the PMNS matrix with the bimaxiaml matrix and the CKM matrix, all
the elements of the PMNS matrix can be expressed by the parameters
of the CKM matrix. The QLC is satisfied perfectly, and Raidal's
relations can be deduced naturally (the deviation from Raidal's
relations is of the order of $\lambda^2$).

{\it Case 2}:
$V_{\mathrm{CKM}}U_{\mathrm{PMNS}}=U_{\mathrm{bimax}}$.

This relation has been pointed out by Giunti and
Tanimoto~\cite{Giunti} and discussed by some other
authors~\cite{Frampton,kang}.
Giunti and Tanimoto~\cite{Giunti} suggested that the deviation of
$U_{\mathrm{PMNS}}$ from the bimaximal mixing matrix is the
CKM-like matrix, and Kang, Kim, and Lee~\cite{kang} got this
relation under the assumptions $Y_u=Y_d^T$, $Y_u=Y_u^T$ in SU(5)
and $Y_{\nu}=Y_u$ in SO(10) grand unified theories.

Repeating the previous process, we can get the PMNS matrix as
\begin{eqnarray}\nonumber
   U_{\mathrm{PMNS}}&=&V_{\mathrm{CKM}}^{\dag}U_{\mathrm{bimax}}\\&=&
\left(
        \begin{array}{ccc}
            \frac{\sqrt{2}}{2} & \frac{\sqrt{2}}{2} & 0 \\
            -\frac{1}{2} & \frac{1}{2} & \frac{\sqrt{2}}{2} \\
            \frac{1}{2} & -\frac{1}{2} & \frac{\sqrt{2}}{2}
        \end{array} \right)+\lambda
\left(
        \begin{array}{ccc}
             \frac{1}{2} & -\frac{1}{2} & -\frac{\sqrt{2}}{2} \\
             \frac{\sqrt{2}}{2} & \frac{\sqrt{2}}{2} & 0\\
             0 & 0 & 0
        \end{array} \right)+\lambda^{2}
\left(
        \begin{array}{ccc}
            -\frac{\sqrt{2}}{4} & -\frac{\sqrt{2}}{4} & 0 \\
            \frac{1}{4}-\frac{1}{2}A & -\frac{1}{4}+\frac{1}{2}A & -\frac{\sqrt{2}}{4}-\frac{\sqrt{2}}{2}A \\
            -\frac{1}{2}A & \frac{1}{2}A & \frac{\sqrt{2}}{2}A
        \end{array} \right)\nonumber\\&&+\lambda^{3}
\left(
        \begin{array}{ccc}
            \frac{1}{2}A(1-\rho+i\eta) & -\frac{1}{2}A(1-\rho+i\eta) & \frac{\sqrt{2}}{2}A(1-\rho+i\eta) \\
            0 & 0 & 0\\
            \frac{\sqrt{2}}{2}A(\rho+i\eta) & \frac{\sqrt{2}}{2}A(\rho+i\eta) & 0
        \end{array} \right)+\cdots.\label{eq.others}
\end{eqnarray}

We can see that the leading term in Eq.~(\ref{eq.others}) is the
bimaximal mixing pattern as that in Eq.~(\ref{eq.3ci}) and
Eq.~(\ref{eq.2ci}). However, from the next-to-leading term, there
are differences between Eq.~(\ref{eq.others}) and
Eq.~(\ref{eq.3ci}) and Eq.~(\ref{eq.2ci}). This indicates that the
degree of the breaking of Raidal's relations
(Eq.~(\ref{eq.raidal})) is larger than that of {\it Case 1}.

Similarly, we can get all the six the trigonometric functions of
the mixing angles of leptons.

From Eq.~(\ref{eq.others}), we have (to the order of $\lambda^3$)
\begin{eqnarray}\nonumber
&&s_1^{\mathrm{PMNS}}=\frac{\sqrt{2}}{2}-\frac{\sqrt{2}}{2}(A+\frac{1}{4})\lambda^2,\nonumber\\
&&c_1^{\mathrm{PMNS}}=\frac{\sqrt{2}}{2}+\frac{\sqrt{2}}{2}(A+\frac{1}{4})\lambda^2,\nonumber\\
&&|s_2^{\mathrm{PMNS}}|=\frac{\sqrt{2}}{2}\lambda\sqrt{[A\lambda^2(1-\rho)-1]^2+(A\lambda^2\eta)^2}=0.68\lambda,\nonumber\\
&&c_2^{\mathrm{PMNS}}=1-0.23\lambda^2,\nonumber\\
&&s_3^{\mathrm{PMNS}}=\frac{\sqrt{2}}{2}-\lambda-\frac{\sqrt{2}}{2}\lambda^2+(A+\frac{1}{2})\lambda^3,\nonumber\\
&&c_3^{\mathrm{PMNS}}=\frac{\sqrt{2}}{2}+\lambda-\frac{\sqrt{2}}{2}\lambda^2-(A+\frac{1}{2})\lambda^3.\label{eq.kang2}
\end{eqnarray}

And we can get the sums of mixing angles of quarks and leptons.
\begin{eqnarray}\nonumber
\sin(\theta_1^{\mathrm{CKM}}+\theta_1^{\mathrm{PMNS}})=\frac{\sqrt{2}}{2}-\frac{\sqrt{2}}{8}\lambda^2,\nonumber
\end{eqnarray}
and thus
\begin{eqnarray}
\theta_1^{\mathrm{CKM}}+\theta_1^{\mathrm{PMNS}}=\frac{\pi}{4}-\frac{1}{4}\lambda^2.\label{eq.kang1}
\end{eqnarray}
And
\begin{eqnarray}\nonumber
&&\sin(\theta_3^{\mathrm{CKM}}+\theta_3^{\mathrm{PMNS}})=\frac{\sqrt{2}}{2}+(\frac{\sqrt{2}}{2}-1)\lambda\\
&&+(1-\frac{3\sqrt{2}}{4})\lambda^2+(A+1-\frac{\sqrt{2}}{2})\lambda^3,\nonumber
\end{eqnarray}
and thus
\begin{eqnarray}\nonumber
&&\theta_1^{\mathrm{CKM}}+\theta_1^{\mathrm{PMNS}}=\frac{\pi}{4}-(\sqrt{2}-1)\lambda\\
&&+(\sqrt{2}-\frac{3}{2})\lambda^2+(\sqrt{2}A+\sqrt{2}-1)\lambda^3.\label{eq.kang3}
\end{eqnarray}

We can see from Eq.~(\ref{eq.kang1}) and Eq.~(\ref{eq.kang3}) that
both of the Raidal's relations break down, and the QLC is broken
to the order of $\lambda$. This breaking has been pointed out by
Minakata and Smirnov~\cite{qlc} and Kang, Kim, and
Lee~\cite{kang}. Comparing with Eq.~(\ref{eq.linan1}) and
Eq.~(\ref{eq.linan2}), we can see that the difference is caused by
the order of the product. If we set
$V_{\mathrm{CKM}}U_{\mathrm{PMNS}}=U_{\mathrm{bimax}}$, the
deviations from Raidal's relations are larger than the results if
we set $U_{\mathrm{PMNS}}V_{\mathrm{CKM}}=U_{\mathrm{bimax}}$.

Also, from Eq.~(\ref{eq.kang2}), we know
$|s_2^{\mathrm{PMNS}}|=0.68\lambda$, so we can get
$|U_{e3}^{\mathrm{PMNS}}|=0.68\lambda$. Substituting the best fit
value $\lambda=0.2243$~\cite{pdg} into it, we have
\begin{eqnarray}\nonumber
|U_{e3}^{\mathrm{PMNS}}|=0.15.
\end{eqnarray}

This value is quite near the upper bound of
$|U_{e3}^{\mathrm{PMNS}}|<0.20$. However, from Eq.~(\ref{eq.s2}),
we know $|s_2^{\mathrm{PMNS}}|=0.48\lambda^2$, so we can get
\begin{eqnarray}\nonumber
|U_{e3}^{\mathrm{PMNS}}|=0.48\lambda^2=0.024.
\end{eqnarray}
We can see that this result is more consistent with the current
experimental upper bound.

From the discussions above, we can see that there are the
non-equivalence of Eq.~(\ref{eq.uvb}) or Eq.~(\ref{eq.vub}) and
Raidal's numerical relations of the mixing angles, which means
that we can not get Raidal's numerical relations of the mixing
angles {\it exactly} from Eq.~(\ref{eq.uvb}) or
Eq.~(\ref{eq.vub}), and vice versa. There are small deviations
from the {\it exact} Raidal's numerical relations of the mixing
angles if we take Eq.~(\ref{eq.uvb}) or Eq.~(\ref{eq.vub}) as
precise results. (For example, see Eq.~(\ref{eq.linan2}) and
Eq.~(\ref{eq.kang3}).)

Furthermore, we find that the product
$U_{\mathrm{PMNS}}V_{\mathrm{CKM}}=U_{\mathrm{bimax}}$ is better
than $V_{\mathrm{CKM}}U_{\mathrm{PMNS}}=U_{\mathrm{bimax}}$ from
the viewpoints of both symmetric and phenomenological
considerations. Of course, if the deviation of the PMNS matrix
from the bimaxiaml mixing matrix is not exactly the CKM matrix,
but is just the CKM-like matrix~\cite{Giunti,Frampton,kang} (i.e.,
the elements of the matrix have the same hierarchy as the
Wolfenstein parametrization, but with not exactly the same
Wolfenstein parameters), Eq.~(\ref{eq.vub}) may still be
satisfied. The two different cases can be further discriminated by
future experiments.

If the relation
$U_{\mathrm{PMNS}}V_{\mathrm{CKM}}=U_{\mathrm{bimax}}$ is
supported by the future experimental data, using
Eq.~(\ref{eq.ckm}) and Eq.~(\ref{eq.pmns}) we have
\begin{eqnarray}\nonumber
U_{\mathrm{PMNS}}V_{\mathrm{CKM}}=U_l^{\dag}U_{\nu}V_u^{\dag}V_d=U_{\mathrm{bimax}}.
\end{eqnarray}

However, we know that $U_l$, $U_{\nu}$, $V_u$ and $V_d$ are not
definite, and we can set $U_l$ and $V_d$ to be the unit matrix by
redefining the quark and lepton fields, and thus we have
\begin{eqnarray}\nonumber
U_{\mathrm{PMNS}}=U_{\nu}, \quad V_{\mathrm{CKM}}=V_u^{\dag},
\end{eqnarray}
and thus
\begin{eqnarray}\nonumber
U_{\mathrm{PMNS}}V_{\mathrm{CKM}}=U_{\nu}V_u^{\dag}=U_{\mathrm{bimax}}.
\end{eqnarray}

So we can find that the relation between the CKM and the PMNS
matrices can be transformed to the relation between $V_u$ and
$U_{\nu}$, and we may regard this the complementarity of quark and
lepton mixing matrices.

\section{Conclusions}

In this paper, we explore the relations between the mixing angles
and mixing matrices of quarks and leptons. For the mixing angles,
with Raidal's relations, we can link the mixing angles of quarks
and leptons in a same framework, and then express their mixing
matrices in a unified way, i.e., we can parameterize the PMNS
matrix with the Wolfenstein parameters of the CKM
matrix~\cite{Li}. With this unified parametrization, we discuss
the relations between the quark and lepton mixing matrices. Both
$V_{\mathrm{CKM}}U_{\mathrm{PMNS}}$ and
$U_{\mathrm{PMNS}}V_{\mathrm{CKM}}$ are calculated in detail, and
we can find that $U_{\mathrm{PMNS}}V_{\mathrm{CKM}}$ is more
closer to the bimaximal mixing matrix than
$V_{\mathrm{CKM}}U_{\mathrm{PMNS}}$.

Similarly, for the relation between the quark and lepton mixing
matrices, if we have
$V_{\mathrm{CKM}}U_{\mathrm{PMNS}}=U_{\mathrm{bimax}}$, we can
find that Raidal's relations will violate, especially the elegant
quark-lepton complementarity (QLC) will break (the degree of
breaking is of order $\lambda$). On the contrary, if we set
$U_{\mathrm{PMNS}}V_{\mathrm{CKM}}=U_{\mathrm{bimax}}$, we can see
that Raidal's relations will hold well to the order of
$\lambda^2$, and the QLC will be a precise relation exactly.
Although $U_{\mathrm{PMNS}}V_{\mathrm{CKM}}=U_{\mathrm{bimax}}$ is
still a phenomenological suggestion, it is consistent with the
experimental data and is supported by the analysis in Sec.~4.
Future experimental discrimination between the two different cases
of $V_{\mathrm{CKM}}U_{\mathrm{PMNS}}=U_{\mathrm{bimax}}$ or
$U_{\mathrm{PMNS}}V_{\mathrm{CKM}}=U_{\mathrm{bimax}}$,
will shed light on our understanding of the relation between the
quark and lepton mixing matrices, and will be also helpful for the
future model construction of the quark and lepton mixing matrices
in a grand unified theory.

{\bf Acknowledgments: } We are grateful to Prof.~Xiao-Gang He for
his stimulating suggestions, and to Feng Huang for some
discussion. This work is partially supported by National Natural
Science Foundation of China (Nos.10025523, 90103007, and
10421003) 
and by the Key Grant Project of Chinese Ministry of Education
(NO.~305001).

\end{onecolumn}


\begin{thebibliography}{99}
\bibitem{ckm}
N.~Cabibbo, Phys. Rev. Lett. {\bf 10}, 531 (1963); M.~Kobayashi
and T.~Maskawa, Prog. Theor. Phys. {\bf 49}, 652 (1973).

\bibitem{pmns}
B.~Pontecorvo, Sov. Phys. JETP {\bf 6}, 429 (1958); Z.~Maki,
M.~Nakagawa, and S.~Sakata, Prog. Theor. Phys. {\bf 28}, 870
(1962).

\bibitem{raidal}
M.~Raidal, Phys. Rev. Lett. {\bf 93}, 161801 (2004).

\bibitem{Li}
N.~Li and B.-Q.~Ma, Phys. Rev. {\bf D 71}, 097301 (2005).

\bibitem{pdg}
Particle Data Group, S.~Eidelman {\it et al.}, Phys. Lett. {\bf B
592}, 1 (2004).

\bibitem{wol}
L.~Wolfenstein, Phys. Rev. Lett. {\bf 51}, 1945 (1983).

\bibitem{garcia}
M.C.~Gonzalez-Garcia, hep-ph/0410030.

\bibitem{qlc}
H.~Minakata and A.Yu.~Smirnov, Phys. Rev. {\bf D 70}, 073009
(2004).

\bibitem{literature}
G.~Altarelli, F.~Feruglio, and I.~Masina, Nucl. Phys. {\bf B 689},
157 (2004); A.~Romanino, Phys. Rev. {\bf D 70}, 013003 (2004);
C.A.~de~S.~Pires, J. Phys. {\bf G 30}, B29 (2004); S.T.~Petcov and
W.~Rodejohann, Phys. Rev. D {\bf 71}, 073002 (2005); 
J.~Ferrandis and S.~Pakvasa, Phys. Lett. {\bf B 603}, 184 (2004).

\bibitem{sch}
J.~Schechter and J.W.F.~Valle, Phys. Rev. {\bf D 22}, 2227 (1980);
S.M.~Bilenky, J.~Hosek, and S.T.~Petcov, Phys. Lett. {\bf B 94},
495 (1980).

\bibitem{Kam}
KamLAND Collaboration, K.~Eguchi {\it et al.}, Phys. Rev. Lett.
{\bf 90}, 021802 (2003).

\bibitem{sno}
SNO Collaboration, S.N.~Ahmed {\it et al.}, Phys. Rev. Lett. {\bf
92}, 181301 (2004).

\bibitem {K2K}
K2K Collaboration, M.H.~Ahn {\it et al.}, Phys. Rev. Lett. {\bf
90}, 041801 (2003).

\bibitem {SUPER}
Super-Kamiokande Collaboration, Y.~Fukuda {\it et al.}, Phys. Rev.
Lett. {\bf 81}, 1562 (1998); Y.~Ashie {\it et al.}, Phys. Rev.
Lett. {\bf 93}, 101801 (2004). C.K.~Jung, C.~McGrew, T.~Kajita,
and T.~Mann, Anna. Rev. Nucl. Part. Sci. {\bf 51}, 451 (2001).

\bibitem{Chz}
CHOOZ Collaboration, M.~Apollonio {\it et al.}, Phys. Lett. {\bf B
420}, 397 (1998); Palo Verde Collaboration, F.~Boehm {\it et al.},
Phys. Rev. Lett. {\bf 84}, 3764 (2000).

\bibitem{Smirnov}
A.Yu.~Smirnov, hep-ph/0402264.

\bibitem{Rodejohann}
W.~Rodejohann, Phys. Rev. {\bf D 69}, 033005 (2004).

\bibitem{li}
N.~Li and B.-Q.~Ma, Phys. Lett. {\bf B 600}, 248 (2004).

\bibitem{Giunti}
C.~Giunti and M.~Tanimoto, Phys. Rev. {\bf D 66}, 053013 (2002);
Phys. Rev. {\bf D 66}, 113006 (2002).

\bibitem{Frampton}
P.H.~Frampton, S.T.~Petcov, and W.~Rodejohann, Nucl. Phys. {\bf B
687}, 31 (2004).

\bibitem{kang}
S.K.~Kang, C.S.~Kim, and J.~Lee, hep-ph/0501029.



\nonfrenchspacing
\end{thebibliography}
\end{document}